\documentclass[twocolumn,aps,prb,preprintnumbers]{revtex4-2}
\usepackage{bbm}
\usepackage[latin9]{inputenc}
\usepackage{amsmath}
\usepackage{amssymb}
\usepackage{graphicx}
\usepackage{mathrsfs}
\usepackage{amsfonts}
\usepackage{amsthm}
\usepackage{color}
\usepackage{txfonts}
\usepackage[colorlinks=true,citecolor=blue,linkcolor=blue,urlcolor=blue,anchorcolor=blue]{hyperref}%
\usepackage{pifont}
\hypersetup{colorlinks=true,citecolor=blue,linkcolor=blue,urlcolor=blue}
\providecommand{\U}[1]{\protect\rule{.1in}{.1in}}
\makeatletter
\@ifundefined{textcolor}{}
{
\definecolor{BLACK}{gray}{0}
\definecolor{WHITE}{gray}{1}
\definecolor{RED}{rgb}{1,0,0}
\definecolor{GREEN}{rgb}{0,1,0}
\definecolor{BLUE}{rgb}{0,0,1}
\definecolor{CYAN}{cmyk}{1,0,0,0}
\definecolor{MAGENTA}{cmyk}{0,1,0,0}
\definecolor{YELLOW}{cmyk}{0,0,1,0}
}

\makeatother

\begin{document}
\title{Frequency comb in twisted magnonic crystals}
\author{Minghao Li}
\author{Zhejunyu Jin}
\author{Zhaozhuo Zeng}
\author{Peng Yan}
\email[Contact author: ]{yan@uestc.edu.cn}
\affiliation{School of Physics and State Key Laboratory of Electronic Thin Films and Integrated Devices, University of Electronic Science and Technology of China, Chengdu 610054, China}

\begin{abstract}
While twisted magnonic crystals (MCs) have recently gained attention for their intriguing linear phenomena, such as magnon flat bands, their nonlinear dynamics---particularly the generation of magnonic frequency combs (MFCs)---have remained largely unexplored. In this work, we demonstrate the creation of MFCs in twisted MCs using two-tone microwave excitation. We find that finite twist angles significantly enhance three-magnon interactions, driven by the non-collinear ground-state magnetic configuration induced by interlayer dipole-dipole interactions. The number of comb teeth exhibits a plateau-like dependence on the twist angle, with the plateau's width and height saturating as the excitation frequency of the propagating magnon mode increases. This behavior reveals an optimal range of twist angles and frequencies for achieving high-quality MFCs with a large number of comb teeth. Our findings deepen the understanding of nonlinear interactions in twisted MCs and highlight their potential for advancing moir\'e-based materials in information processing and high-precision metrology.
\end{abstract}
\maketitle

\section{Introduction}\label{section:A}
Frequency combs consist of a series of discrete, equally spaced spectral lines in the frequency domain. Initially proposed and extensively studied in optical systems, frequency combs serve as a bridge between microwave and optical frequencies \cite{David2000,Udem2002,Fortier2019}. 
Their exceptional precision and stability have led to a wide range of applications, including atomic clocks \cite{Ludlow2019}, precise wavelength calibration \cite{Murphy2007}, and broadband spectroscopy of atoms and molecules \cite{Picque2019}. Drawing inspiration from the success of optical frequency combs, researchers have extended the concept to magnetic systems, giving rise to magnonic frequency combs (MFCs). MFCs can be generated through several mechanisms, such as (\romannumeral1) Nonlinear interactions among magnons \cite{Hula2022,Rao2023,Rao2024}, (\romannumeral2) Nonlinear interactions between magnons and intrinsic modes of spin textures \cite{Wang2021,Zhou2021,Wang2022,Liu2023,Jin2023,Yao2023,Liu2024,Shen2024,Liang2024,Zhang2024,LiuXJ2024}, (\romannumeral3) Magnon-phonon interactions \cite{Xu2023}, and (\romannumeral4) Magnon-photon interactions \cite{Xu2020,Liu2022,Lu2024}.
Despite the variety of methods available for generating MFCs, efficiently controlling their properties---such as the spacing or the number of comb teeth---remains a significant challenge.

Moir\'e superlattices, formed by stacking periodic lattices with a relative twist or lattice constant mismatch, are characterized by the emergence of flat bands at specific ``magic" angles, as first predicted in twisted bilayer graphene \cite{Bistritzer2011}.
The experimental discovery of unconventional superconductivity and correlated insulating states in such systems \cite{Cao2018,CaoY2018} has sparked widespread research into their novel physics. Moir\'e superlattices realized in diverse platforms ranging from twisted atomic layers to artificial photonic \cite{Dong2021} and phononic crystals \cite{Deng2020}, exhibit a variety of exotic phenomena, including the fractional quantum anomalous Hall effect \cite{Park2023}, moir\'e excitons \cite{Seyler2019,Jin2019}, effective gauge field \cite{WangWH2020,Duan2023} and moir\'e quasi bound states \cite{Huang2022}, etc. 
In addition, moir\'e superlattices also display intriguing nonlinear physics including but not limited to nonlinear anomalous Hall effect, second-harmonic generation, high-harmonic generation and optical solitons \cite{Huang2022,Zhang2022,He2022,Du2017,Yao2021,Ikeda2020,Fu2020}. 
Therefore, moir\'e superlattices offer a novel controllable platform for fundamental research and technological innovation. 
In the field of magnetism, studies have explored topological flat bands \cite{Cheng2020}, spin textures \cite{Tong2018,Kim2024}, and domain wall magnons \cite{WangC2020} in magnetic moir\'e superlattices.
Recently, twisted magnonic crystals (MCs) have been proposed, showcasing unique magnon band structures \cite{Chen2022,Wang2023,Chen2024}.
However, the nonlinear physics of twisted MCs remains largely unexplored.

In this paper, we report the generation of MFCs in MCs, a novel system that exploits the unique properties of moir\'e superlattices to enhance nonlinear magnon interactions. Micromagnetic simulations reveal that twisting the MCs disrupts the uniform magnetization distribution of the ground state, breaking the translational symmetry typically observed in untwisted systems. This non-uniformity, driven by interlayer dipole-dipole interactions, significantly strengthens the nonlinear interactions between magnons, consistent with findings in other symmetry-broken magnetic systems \cite{Wang2021}. 
Furthermore, the intricate interlayer couplings in twisted MCs introduce additional magnon modes within the original band gap \cite{Wang2023}, enriching the system's spectral response and facilitating frequency comb generation. To excite the MFC in twisted MCs, we apply a two-tone microwave drive, utilizing two distinct frequencies: one corresponding to a propagating magnon mode ($\omega_1$) and the other to the Kittel mode ($\omega_2$), which is the uniform precession resonance inherent to ferromagnetic materials. This two-tone approach is essential for two key reasons:
(i) A single-frequency microwave drive requires an amplitude threshold so high that it risks destabilizing the ground state, rendering it impractical for stable operation. (ii) When only a single frequency is used, it non-resonantly excites additional magnon modes beyond the Kittel mode, which interfere with the targeted nonlinear processes and diminish the efficiency of MFC generation. By contrast, the two-tone drive selectively enhances the desired interactions at lower amplitudes, ensuring efficient and controlled comb production. The twist angle, a critical parameter in moir\'e superlattices, significantly influences the MFC characteristics, particularly in tandem with the excitation frequency of the propagating magnon mode ($\omega_1$). Our simulations indicate that the number of comb teeth as a function of the twist angle exhibits a plateau-like trend: within a specific range of twist angles---approximately $5^{\circ}$ to $15^{\circ}$---the comb maintains a high number of teeth (up to 20), and this plateau broadens and stabilizes as $\omega_1$ increases. This behavior highlights the tunability of twisted MCs, where both the geometric twist angle and the dynamic excitation frequency can be adjusted to optimize the frequency comb's properties.
Finally, twisted MCs can be fabricated using state-of-the-art micro- and nanofabrication techniques \cite{Wang2023, Mao2021}, enabling their integration into on-chip magnonic devices with excellent tunability. This capability positions twisted MCs as an ideal platform for advanced applications, such as high-precision metrology, signal processing, and quantum information technologies, where frequency combs play a transformative role in analogous optical systems.

The paper is organized as follows. In Sec. \ref{section:B}, we develop a theoretical framework to model the nonlinear interactions between propagating magnons and the Kittel mode within the distinctive setting of twisted MCs. Section \ref{section:C} presents detailed micromagnetic simulations that validate our theoretical predictions and illustrate the behavior of MFCs under various conditions in twisted MCs. Finally, Section \ref{section:D} summarizes our findings and explores their implications for future research and potential applications of twisted MCs in advanced technologies.

\section{Theoretical Model}\label{section:B}
\begin{figure}[t]
    \centering
    \includegraphics[width=0.48\textwidth]{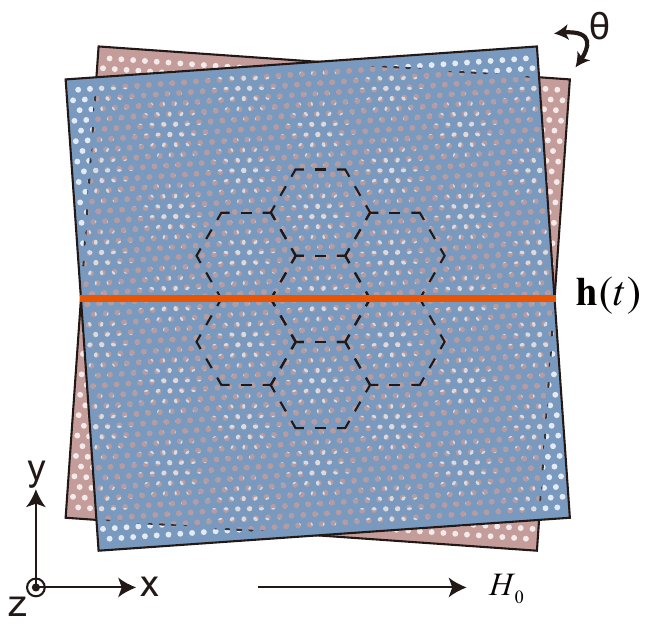}
    \caption{Schematic of the twisted magnonic crystal. Two antidot MC layers are twisted relative to each other by an angle $\theta$, forming a moir\'e superlattice (dashed black lines). A bias magnetic field $H_0$ is applied along the $x$-direction. Microwave fields ${\bf h}(t)$ are applied in the orange stripe region to excite magnons propagating along the $y$-direction.}
    \label{Fig_scheme}
\end{figure}

We consider a system comprising two stacked layers of MCs, each featuring a triangular antidot lattice. An antidot lattice consists of a magnetic film with a periodic array of holes (antidots) where the magnetic material is absent. In this setup, one layer is twisted clockwise by $\theta/2$, and the other counterclockwise by $\theta/2$, resulting in a total twist angle of $\theta$ between the layers, as shown in Fig. \ref{Fig_scheme}. A uniform bias magnetic field is applied along the $x$-direction. The Hamiltonian of the system is expressed as the sum of intralayer, interlayer, and Zeeman interactions
\begin{equation}\label{Eq_hamiltonian1}
    \mathcal{H}=\mathcal{H}_{\text{intra}}+\mathcal{H}_{\text{inter}}+\mathcal{H}_{\text{ext}},
\end{equation}
where $\mathcal{H}_{\text{intra}}$ describes interactions within each layer, $\mathcal{H}_{\text{inter}}$ accounts for interactions between the layers, and $\mathcal{H}_{\text{ext}}$ represents the coupling to the external magnetic field. These terms are given by 
\begin{widetext}
    \begin{subequations}\label{Eq_hamiltonian2}
    \begin{eqnarray}
        \mathcal{H}_{\text{intra}}&=&\sum_{l=t,b}\left[d_l \int A_{ex}({\bf r})(\nabla {\bf m}_l)^2d^2r-\frac{1}{2}\mu_0 d_l^2\int M_s({\bf r}){\bf m}_l({\bf r})\cdot {\tensor N}({\bf r},{\bf r}')\cdot{\bf m}_l({\bf r}')M_s({\bf r}')\,d^2rd^2r'\right], \\        
        \mathcal{H}_{\text{inter}}&=&-\int J_\perp({\bf r}){\bf m}_t({\bf r})\cdot{\bf m}_b({\bf r})d^2r-\sum_{\substack{l,l'=t,b\\l\neq l'}}\frac{1}{2}\mu_0d_l^2\int M_s({\bf r}){\bf m}_l({\bf r})\cdot {\tensor N_\perp}({\bf r},{\bf r}')\cdot{\bf m}_{l'}({\bf r}')M_s({\bf r}')\,d^2rd^2r', \\     
        {\mathcal H}_{\text{ext}}&=&-\sum_{l=t,b}d_l\mu_0\int {\bf H}\cdot{\bf m}_l({\bf r})M_s({\bf r}) d^2r.
    \end{eqnarray}
    \end{subequations}
\end{widetext}

Here ${\bf m}_l({\bf r})$ is the normalized magnetization vector at position ${\bf r} = (x, y)$
 in layer $l$, where $l = t$ denotes the top layer and $l = b$ the bottom layer. The thickness of each layer is $d_l$, the interlayer separation is $d_\perp$, and $\mu_0$ is the vacuum permeability. The intralayer exchange coupling is $A_{\text{ex}}({\bf r})$, and the interlayer exchange coupling is $J_\perp({\bf r})$. The demagnetization tensors ${\tensor N}({\bf r},{\bf r}')$ and ${\tensor N}_\perp({\bf r},{\bf r}')$ describe the intralayer and interlayer dipole-dipole interactions (DDIs), respectively, and are defined as (see Ref. \cite{newell1993})
\begin{equation}\label{Eq_demag_tensor}
    {\tensor N}_{(\perp)}({\bf r},{\bf r}')=\frac{1}{4\pi|\delta|^5}
    \begin{pmatrix}
	3\delta_x^2-|\delta|^2 & 3\delta_x\delta_y & 3\delta_x\delta_z \\
	3\delta_y\delta_x & 3\delta_y^2-|\delta|^2 & 3\delta_y\delta_z \\
	3\delta_z\delta_x & 3\delta_z\delta_y & 3\delta_z^2-|\delta|^2 \\
    \end{pmatrix},
\end{equation}
where ${\boldsymbol \delta} = {\bf r}-{\bf r}'$ for ${\tensor N}({\bf r},{\bf r}')$, ${\boldsymbol \delta} = {\bf r}-{\bf r}'-{\bf \Delta}_{ll'}$ for ${\tensor N}_\perp({\bf r},{\bf r}')$, and ${\bf \Delta}_{ll'}$ is the normal vector pointing from layer $l$ to layer $l'$ with length $|{\bf \Delta}_{ll'}|=d_\perp$. ${\bf H}$ represents the external magnetic field. The saturation magnetization $M_s$ and the exchange coupling $A_{ex}$ are spatially modulated by the antidots, 

\begin{equation}\label{Eq_antidots}
    \begin{cases}
    M_s({\bf r})=M_sf({\bf r})\\
    A_{ex}({\bf r})=A_{ex}f({\bf r})
    \end{cases}\text{\  with }f({\bf r})= 
    \begin{cases}
    1, & {\bf r}\in \text{antidots} \\
    0, & \text{otherwise}
    \end{cases}.
\end{equation}

To study nonlinear magnon-magnon interactions, we quantize the magnetization vectors using the Holstein-Primakoff (HP) transformation \cite{Holstein1940}, which expresses the magnetization in terms of magnon creation and annihilation operators. The Hamiltonian is then expanded in powers of $1/\sqrt{S}$, where $S$ is the spin magnitude, and written as $\mathcal{H}=\mathcal{H}^{(0)}+\mathcal{H}^{(2)}+\mathcal{H}^{(3)}+\mathcal{H}^{(4)}+\cdots$, where $\mathcal{H}^{(0)}$, $\mathcal{H}^{(2)}$, $\mathcal{H}^{(3)}$, and $\mathcal{H}^{(4)}$ correspond to the ground state energy, quadratic (harmonic) magnon dispersion, three- and four-magnon interactions, respectively, with higher-order terms typically neglected for simplicity. In untwisted MCs, the bias magnetic field and exchange interactions generally stabilize a uniform ferromagnetic ground state. Due to the symmetry of the uniform ground state, the exchange interaction primarily generates even-order terms in the Hamiltonian, e.g., $\mathcal{H}^{(0)}$, $\mathcal{H}^{(2)}$, and $\mathcal{H}^{(4)}$, while odd-order terms like $\mathcal{H}^{(3)}$ vanish. As a result, three-magnon interactions are typically weak in untwisted MCs. In contrast, twisted MCs introduce interlayer DDI and a complex moir\'e geometry, which disrupt the uniformity of the static magnetization distribution. This non-uniformity breaks the symmetry that suppresses odd-order terms, significantly enhancing the three-magnon interaction term $\mathcal{H}^{(3)}$. In this case, the exchange interaction plays a key role in driving the enhanced $\mathcal{H}^{(3)}$. A simplified example of this effect in a two-spin system is provided in Appendix \ref{section:two-spin_sys}. 

\begin{figure}[t]
    \centering
    \includegraphics[width=0.48\textwidth]{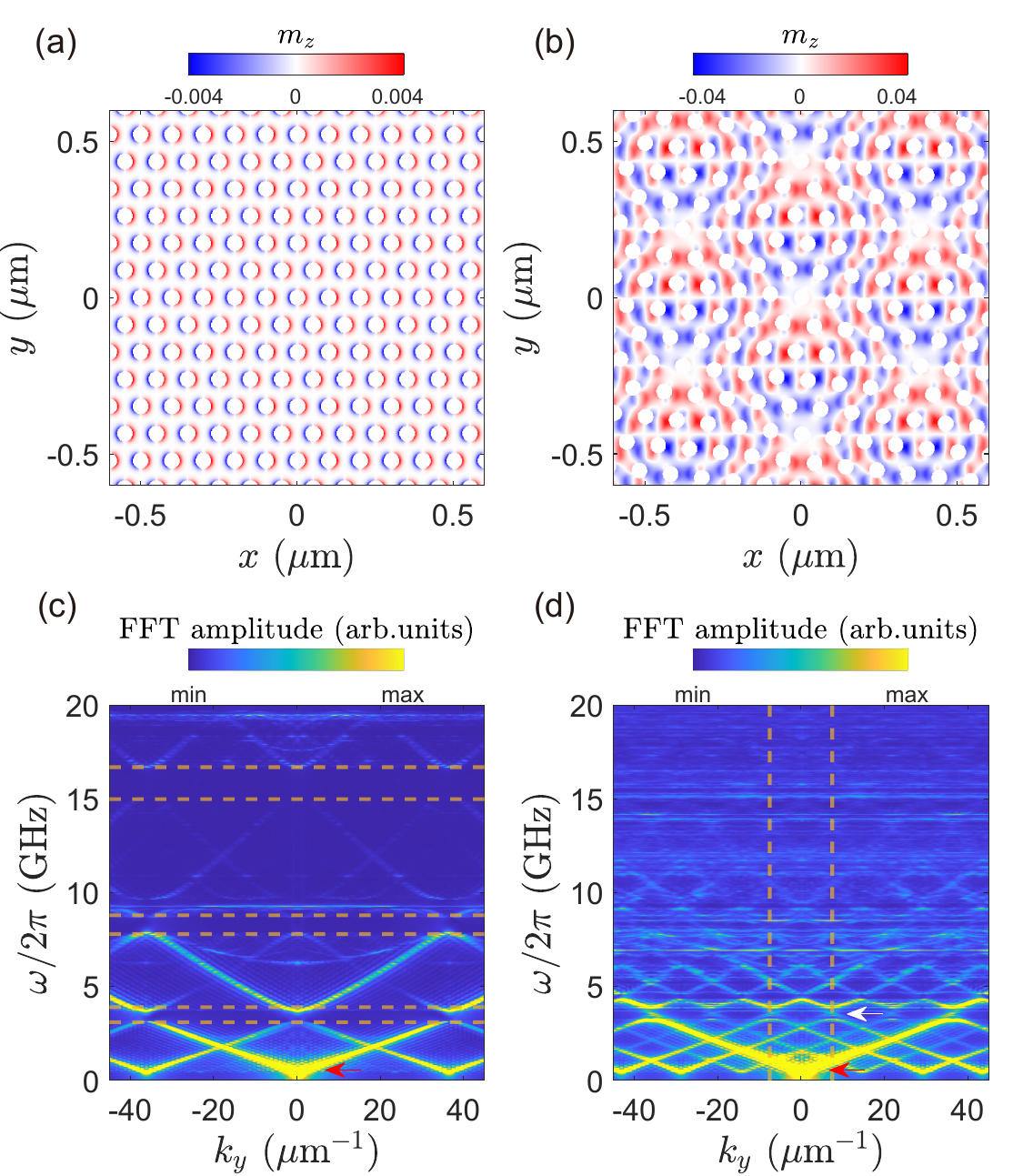}
    \caption{(a) Spatial distribution of the $z$-component of the normalized static magnetization for the untwisted MC. (b) Same as (a) but for the $13^\circ$-twisted MC. (c) Magnon dispersion for the untwisted MC, with dashed orange lines indicating band gaps at 3.1-3.9 GHz, 7.8-8.8 GHz, and 15.0-16.7 GHz. The red arrow at the bottom indicates the Kittel mode at 0.5 GHz. (d) Magnon dispersion for the $13^\circ$-twisted MC, with dashed orange lines denoting the folded first Brillouin zone due to the moir\'e superlattice.}
    \label{Fig_ground_state}
\end{figure}

We investigate the ground states of both untwisted and twisted MCs using micromagnetic simulations with magnetic parameters corresponding to yttrium iron garnet (YIG) films. Details of the simulation setup are provided in Sec. \ref{section:C}. In the untwisted MC, the out-of-plane component of the normalized magnetization, $m_z$, is primarily localized at the edges of the antidots, exhibiting a maximum magnitude of approximately $4 \times 10^{-3}$, as illustrated in Fig. \ref{Fig_ground_state}(a). In contrast, for the twisted MC, the spatial distribution of $m_z$ forms a periodic pattern modulated by the moir\'e superlattice, with a maximum magnitude of approximately $4 \times 10^{-2}$, an order of magnitude larger than in the untwisted case, as shown in Fig. \ref{Fig_ground_state}(b). This significant alteration in the ground state configuration suggests that nonlinear interactions can be enhanced in the twisted MC.

The magnon dispersion relation $\omega(\mathbf{k})$ is determined by the second-order Hamiltonian $\mathcal{H}^{(2)}$. A widely used approach to calculate $\omega(\mathbf{k})$ is the plane-wave expansion method. However, when applied to antidot MCs, this method assumes magnetization pinning at the antidot edges \cite{Klos2012}, a condition not implemented in our micromagnetic simulations. As a result, we present numerically computed dispersion relations for untwisted and twisted MCs in Figs. \ref{Fig_ground_state}(c) and \ref{Fig_ground_state}(d), respectively. The lowest-frequency mode in both cases, marked by red arrows in Figs. \ref{Fig_ground_state}(c) and \ref{Fig_ground_state}(d), is identified as the uniform Kittel mode. For a thin ferromagnetic film, the Kittel mode frequency is expressed as
\begin{equation}\label{Eq_Kittel}
    \omega_l/2\pi=\gamma\mu_0\sqrt{H_0(H_0+M_{s})},
\end{equation}
where $\gamma = 28 \, \text{GHz/T}$ is the gyromagnetic ratio and $H_0$ is the external bias magnetic field applied along the $x$-direction. This formula can be derived from the Hamiltonian in Eq. (\ref{Eq_hamiltonian2}) by assuming uniform precession of the magnetization vectors and approximating the demagnetization tensor for a thin film as
\begin{equation}\label{Eq_N_thin_film}
        {\tensor N}_{\text{film}}({\bf r},{\bf r}')=\frac{1}{V}\delta({\bf r}-{\bf r}')
    \begin{pmatrix}
	0 & 0 & 0 \\
	0 & 0 & 0 \\
	0 & 0 & 1 \\
    \end{pmatrix}
\end{equation}
where $V$ is the total volume of the film and $\delta(\mathbf{r} - \mathbf{r}')$ is the Dirac delta function. The Kittel mode, being the lowest-frequency magnon mode excitable with the highest excitation efficiency in our simulations [as indicated by the brightest color marked by red arrows in Figs. \ref{Fig_ground_state}(c) and \ref{Fig_ground_state}(d)], is particularly significant and serves as the frequency spacing for the MFC. In the twisted MC, the moir\'e superlattice induces a band folding effect, as indicated by the folded first Brillouin zone outlined with dashed orange lines in Fig. \ref{Fig_ground_state}(d). This modification of the magnon band structure may facilitate novel magnon interactions and enhance the system's nonlinear properties.

To investigate the physical mechanisms driving the formation of a MFC, we analyze a simplified model involving four distinct magnon modes: the incident (propagating) magnon mode $a_k$, the Kittel mode $a_l$, the confluence mode $a_p$, and the splitting mode $a_q$. Here, the Kittel mode $a_l$ represents the uniform precession mode of the ferromagnetic material, while $a_k$ is a propagating magnon with a specific wavevector. The modes $a_p$ and $a_q$ are secondary magnon states excited through nonlinear three-magnon interactions, specifically confluence and splitting processes, respectively. The Hamiltonian describing these nonlinear interactions, up to third order, is expressed as
 \begin{equation}\label{Eq_hamiltonian3}
    \begin{aligned}
        \mathcal{H}=&\sum_{\nu=k,l,p,q}{\omega_\nu a_\nu^{\dagger}a_\nu}+g_p(a_ka_la_p^\dagger+{\rm H.c.})+g_q(a_ka_l^\dagger a_q^\dagger+{\rm H.c.})\\
        &+\lambda_1 h_1(a_ke^{i\omega_1t}+a_k^\dagger e^{-i\omega_1t})+\lambda_2 h_2(a_le^{i\omega_2t}+a_l^\dagger e^{-i\omega_2t}).
    \end{aligned}
\end{equation}
Here $\omega_\nu$ (for $\nu = k, l, p, q$) denotes the natural frequency of each magnon mode. $g_p$ and $g_q$ are the coupling strengths for the three-magnon confluence and splitting processes, respectively, with ``H.c." indicating the Hermitian conjugate. The terms $\lambda_1 h_1 (a_k e^{i \omega_1 t} + a_k^\dagger e^{-i \omega_1 t})$ and $\lambda_2 h_2 (a_l e^{i \omega_2 t} + a_l^\dagger e^{-i \omega_2 t})$ represent the coherent excitation of modes $a_k$ and $a_l$ by external microwave fields oscillating at frequencies $\omega_1$ and $\omega_2$, respectively. The coefficients $\lambda_1$ and $\lambda_2$ quantify the excitation efficiency, determined by the Fourier components of the microwave fields' spatial distribution. The nonlinear terms in the Hamiltonian describe two key processes: three-magnon confluence $g_p (a_k a_l a_p^\dagger + \text{H.c.})$, a process where two magnons from modes $a_k$ and $a_l$ combine to create a magnon in mode $a_p$, satisfying the frequency condition $\omega_p \approx \omega_k + \omega_l$, and three-magnon splitting $g_q (a_k a_l^\dagger a_q^\dagger + \text{H.c.})$, a process where a magnon in mode $a_k$ splits into magnons in modes $a_l$ and $a_q$, with $\omega_k \approx \omega_l + \omega_q$. Using this Hamiltonian, the Heisenberg equations of motion for the magnon operators are derived, incorporating phenomenological damping terms to account for energy dissipation
 \begin{equation}\label{Eq_Heisenberg}
    \begin{aligned}
        i\frac{{\rm d}a_k}{{\rm d}t}&=(\omega_k-i\alpha_k\omega_k)a_k+g_q a_l a_q+g_p a_l^\dagger a_p+\lambda_1 h_1e^{-i\omega_1t},\\
        i\frac{{\rm d}a_l}{{\rm d}t}&=(\omega_l-i\alpha_l\omega_l)a_l+g_q a_k a_q^\dagger+g_p a_k^\dagger a_p+\lambda_2 h_2e^{-i\omega_2t},\\
        i\frac{{\rm d}a_p}{{\rm d}t}&=(\omega_p-i\alpha_p\omega_p)a_p+g_p a_k a_l,\\
        i\frac{{\rm d}a_q}{{\rm d}t}&=(\omega_q-i\alpha_q\omega_q)a_q+g_q a_k a_l^\dagger,\\
    \end{aligned}
\end{equation}
where $\alpha_\nu$ (for $\nu = k, l, p, q$) are the effective Gilbert damping constants, introduced to model the decay of each mode due to intrinsic losses in the system. Once the confluence mode $a_p$ and splitting mode $a_q$ are excited via these nonlinear interactions, they can further couple with the Kittel mode $a_l$. This triggers a cascade of three-magnon processes, where each interaction generates new frequency components---for instance, $\omega_p = \omega_k + \omega_l$ from confluence or $\omega_q = \omega_k - \omega_l$ from splitting. These newly created frequencies can participate in additional nonlinear interactions, leading to a series of equidistant spectral lines. This cascading mechanism ultimately results in the formation of the MFC, a hallmark of rich nonlinear dynamics in magnonic systems.

\section{Micromagnetic Simulations} \label{section:C}
To gain a comprehensive understanding of the physical processes described above, we perform full micromagnetic simulations using MuMax3, an open-source GPU-accelerated micromagnetic simulation program \cite{Vansteenkiste2014}. The simulated system consists of two stacked MCs, each featuring a 100-nm-period triangular antidot lattice with antidot diameters of 50 nm. This structure is achievable using current micro- and nanofabrication techniques \cite{Wang2023,Mao2021}. Each layer has a thickness of 5 nm, and the total simulation domain measures $3.5 \, \mu\text{m} \times 3.5 \, \mu\text{m} \times 10 \, \text{nm}$, discretized into a cell size of $5 \, \text{nm} \times 5 \, \text{nm} \times 5 \, \text{nm}$. The material parameters are selected to match those of YIG, a widely used low-damping magnetic material: saturation magnetization $M_s = 140 \, \text{kA/m}$, exchange stiffness $A_{\text{ex}} = 3.7 \times 10^{-12} \, \text{J/m}$, interlayer exchange constant $J_\perp=2A_{\text{ex}}/d_\perp$ and Gilbert damping $\alpha = 10^{-4}$. To prevent magnon reflections at the boundaries, absorbing boundary conditions are implemented \cite{Venkat2018}. A static bias magnetic field of 2 mT is applied along the $x$-direction to stabilize the magnetic ground state.

To characterize the magnonic band structure, we apply a sinc-function magnetic field $\mathbf{h}(t) = h_0 \text{sinc}(\omega_c t) \hat{\mathbf{y}}$, with an amplitude of $\mu_0 h_0 = 100 \, \text{mT}$ and a cutoff frequency of $\omega_c / 2\pi = 20 \, \text{GHz}$, to a 20-nm-wide stripe region [orange line in Fig. \ref{Fig_scheme}] in the untwisted MC, $\theta = 0^\circ$. By performing a fast Fourier transform (FFT) on the dynamic magnetization component $\delta m_y$, we obtain the magnonic dispersion relation, as shown in Fig. \ref{Fig_ground_state}(c). The lowest-frequency mode $\omega_l / 2\pi = 0.5 \, \text{GHz}$ at $k = 0$ corresponds to the uniform Kittel mode. Distinct band gaps appear at 3.0-3.8 GHz, 7.8-8.8 GHz, and 15.0-16.7 GHz [dashed orange lines in Fig. \ref{Fig_ground_state}(c)], which impede MFC formation, as frequencies within these gaps are difficult to excite. For a twisted MC with $\theta = 13^\circ$, the dispersion relation, shown in Fig. \ref{Fig_ground_state}(d), is significantly modified due to enhanced interlayer interactions induced by the moir\'e superlattice. The original band gaps' boundaries become less pronounced, and new magnon modes emerge within these gaps [white arrow in Fig. \ref{Fig_ground_state}(d)]. These modes, identified as moir\'e edge or cavity modes within mini-flatbands, or ordinary propagating modes, depend on the specific combination of twist angle $\theta$ and bias magnetic field $H_0$ \cite{Wang2023}. These additional modes facilitate MFC generation by populating the previously inaccessible band gap frequencies. Notably, the Kittel mode frequency remains unchanged at 0.5 GHz in both untwisted and twisted configurations. 

\begin{figure} [h]
    \centering
    \includegraphics[width=0.48\textwidth]{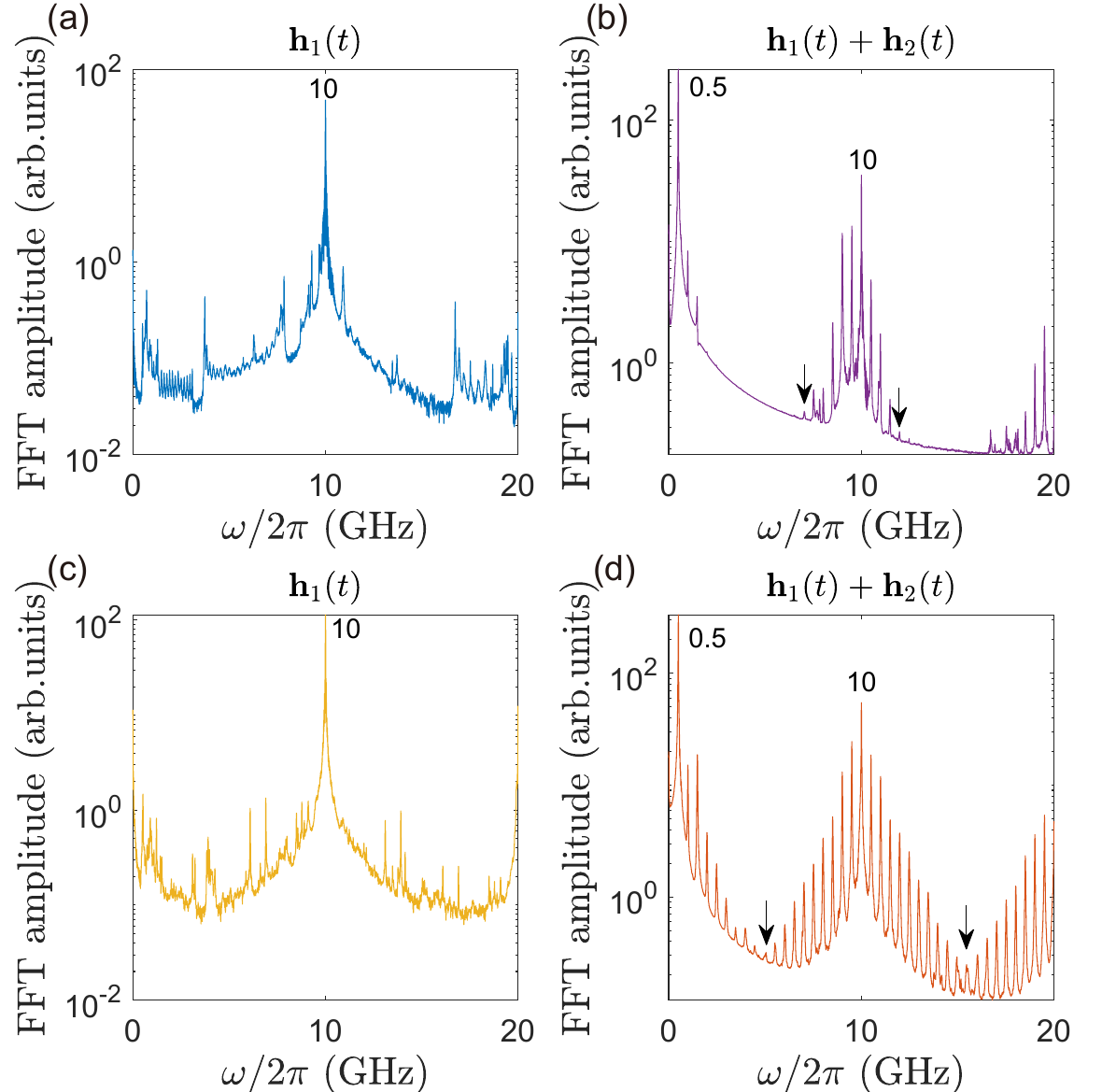}
    \caption{(a) (c) The magnon spectra of the upper layer of untwisted MC and  $13^\circ$ twisted MC when only the exciting field $\mu_0h_1=60\,{\rm mT}$ is applied, respectively. (b) (d) The magnon spectrum of the upper layer of untwisted MC and  $14^\circ$ twisted MC when two exciting fields $\mu_0h_1=60\,{\rm mT}$ and $\mu_0h_2=5\,{\rm mT}$ are applied, respectively.}
    \label{Fig_two_tone}
\end{figure}

We attempt to generate a MFC by applying a single-frequency magnetic field $\mathbf{h}_1(t) = h_1 \sin(\omega_1 t) \hat{\mathbf{y}}$, with $\omega_1 / 2\pi = 10 \, \text{GHz}$ and $\mu_0 h_1 = 60 \, \text{mT}$, to the stripe region shown in Fig. \ref{Fig_scheme}. The magnon spectrum is obtained by performing a FFT on the time-domain data of the dynamic magnetization component $\delta m_y$ for each simulation cell, followed by averaging the resulting spectra across the entire film. However, as shown in Figs. \ref{Fig_two_tone}(a) and \ref{Fig_two_tone}(c) for the untwisted and $13^\circ$-twisted MCs, respectively, no evidence of an MFC is observed in either configuration. This absence suggests that the cascaded three-magnon processes, which are critical for MFC formation, do not occur under single-frequency excitation. The likely reason is the insufficient strength of nonlinear interactions when only one magnon mode is driven, leaving the Kittel mode unexcited and preventing the necessary mode coupling.

To generate the MFC, we apply an additional microwave field $\mathbf{h}_2(t) = h_2 \sin(\omega_2 t) \hat{\mathbf{y}}$ with $\omega_2 = \omega_l = 0.5 \, \text{GHz}$, and amplitude $\mu_0 h_2 = 5 \, \text{mT}$. The magnon spectra for the untwisted and $13^\circ$-twisted MCs are presented in Figs. \ref{Fig_two_tone}(b) and \ref{Fig_two_tone}(d), respectively. In both configurations, three distinct sets of magnon signals are observed around 0.5 GHz, 10 GHz, and 20 GHz:
The signals around 0.5 GHz correspond to frequency multiplications of the Kittel mode, i.e., $m \omega_l$, where $m$ is a positive integer. The signals around 10 GHz represent MFC modes generated through nonlinear interactions between the incident magnons (at frequency $\omega_1$) and the Kittel mode, specifically $\omega_1 \pm m \omega_l$. The signals around 20 GHz are MFC modes arising from the interaction between the second harmonic generation (SHG) of the incident magnons and the Kittel mode, i.e., $2 \omega_1 \pm m \omega_l$. Notably, the MFC in the $13^\circ$-twisted MC exhibits a significantly larger number of comb teeth, up to 22, in stark contrast to the untwisted MC. The diminished MFC quality in the untwisted case is attributed to the presence of band gaps, which are challenging to avoid for $0 < \omega_1 / 2\pi < 20 \, \text{GHz}$, and relatively weak nonlinear interaction strengths. 
It is worthwhile discussing three-magnon process associated with the (mini) flatband, which is the most distinctive characteristic of a moir\'e system. Three-magnon process occurring merely within the flatband is forbidden by the conservation law of energy and momentum, while the scattering of magnon modes from other magnon bands into the flatband, or the reverse process is theoretically possible. Nevertheless, our numerical simulations reveal that MFCs involving flatband modes do not demonstrate superior efficiency compared to those utilizing propagating mode and Kittel mode.

\begin{figure*}[bth]
    \centering
    \includegraphics[width=0.96\textwidth]{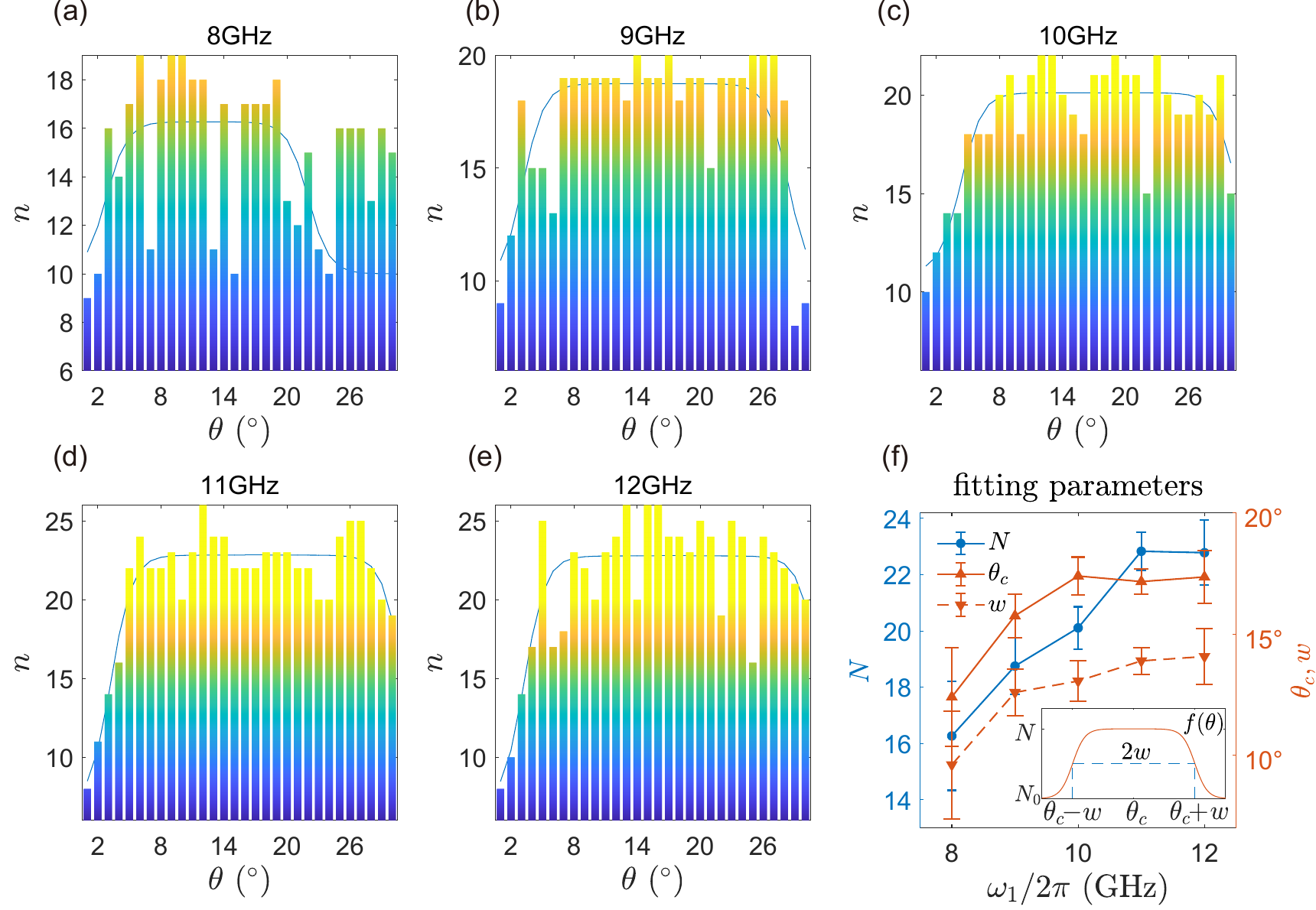}
    \caption{(a)-(e) comb teeth as a function of twist angle $\theta$ for excitation frequencies $\omega_1/2\pi=8,9,10,11$ and $12\,\rm GHz$. The blue solid lines represent fitting curves to the data. (f) Dependence of the fitting parameters on the excitation frequency $\omega_1$. Error bars indicate the 95\% confidence bounds. The inset illustrates the fitting function $f(\theta)$ with labeled parameters $N$, $\theta_c$, $w$, and $N_0$.}
    \label{Fig_angle}
\end{figure*}

Next, we investigate the dependence of the MFC on the twist angle $\theta$ for various excitation frequencies $\omega_1$. We perform micromagnetic simulations on twisted MCs with twist angles $\theta$ ranging from $1^\circ$ to $30^\circ$, covering the full symmetry range of the triangular lattice. The excitation frequencies are set to $\omega_1 / 2\pi =$ 8, 9, 10, 11, and 12 GHz, with fixed parameters $\mu_0 h_1 = 60 \, \text{mT}$, $\mu_0 h_2 = 5 \, \text{mT}$, and $\omega_2 / 2\pi = 0.5 \, \text{GHz}$. To evaluate the MFC quality, we count the number of comb teeth $n$ generated by the incident magnon at frequency $\omega_1$. Specifically, $n$ represents the number of spectral peaks between the lowest peaks on two sides of $\omega_1 / 2\pi$, as shown between the black arrows in Figs. \ref{Fig_two_tone}(b) and \ref{Fig_two_tone}(d) and plotted in Figs. \ref{Fig_angle}(a)-(e). For excitation frequencies other than 8 GHz, the number of teeth $n$ as a function of $\theta$ follows a plateau-like pattern: it increases at small angles, stabilizes within a specific angular range, and decreases at larger angles. The exception at 8 GHz arises because it lies within the original band gap of 7.8-8.8 GHz, making the MFC highly sensitive to the twist angle. To quantify this behavior, we fit the data to a plateau-like function 
\begin{equation}\label{Eq_fit_angle}
    f(\theta)=N_1\sigma(\theta-\theta_c+w)\sigma\big[-(\theta-\theta_c)+w\big]+N_0,
\end{equation}
where $\sigma(\theta)$ is the sigmoid function
\begin{eqnarray}\label{Eq_sigmoid}
    \sigma(\theta)=\frac{1}{1+e^{-\theta}}.
\end{eqnarray}
Here, $N_1$ is the plateau height, $\theta_c$ is the central angle, $w$ is the half-width of the plateau, and $N_0$ is the number of teeth for the untwisted MC. Thus, the maximum number of teeth is $N = N_1 + N_0$. The fitting curves are shown as solid blue lines in Figs. \ref{Fig_angle}(a)-(e), with extracted parameters plotted in Fig. \ref{Fig_angle}(f). The results reveal clear trends: the maximum number of teeth $N$ increases with $\omega_1$ from 8 to 11 GHz, saturating at approximately 23 teeth beyond 11 GHz. The central angle $\theta_c$ and half-width $w$ increase with $\omega_1$ up to 10 GHz, stabilizing at $17^\circ$ and $13^\circ$, respectively. These findings indicate that MFCs achieve high quality, with more than 20 teeth, for excitation frequencies $10 < \omega_1 / 2\pi < 12 \, \text{GHz}$ and twist angles $10^\circ < \theta < 25^\circ$.

\begin{figure}[tbh]
    \centering
    \includegraphics[width=0.48\textwidth]{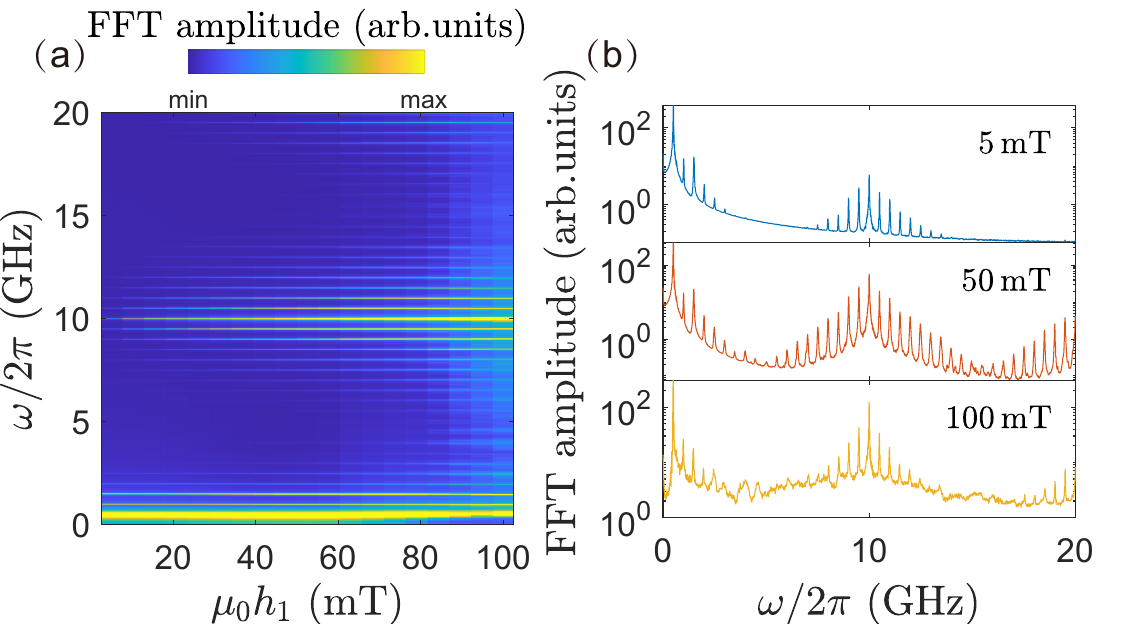}
    \caption{(a) Magnon spectra as a function of the driving field amplitude $\mu_0 h_1$ for a twisted magnonic crystal with a fixed twist angle $\theta = 13^\circ$ and excitation frequencies $\omega_1 / 2\pi = 10 \, \text{GHz}$, $\omega_2 / 2\pi = 0.5 \, \text{GHz}$. (b) FFT spectra at driving field amplitudes $\mu_0 h_1 = 5, 50$, and $100 \, \text{mT}$, illustrating the evolution of the magnonic frequency comb with increasing field strength.}
    \label{Fig_h1}
\end{figure}

Since the twisted MC with a twist angle of $\theta = 13^\circ$ exhibits one of the highest numbers of comb lines at a driving frequency of $\omega_1 / 2\pi = 10 \, \text{GHz}$, we focus primarily on this configuration. Figure \ref{Fig_h1} presents the FFT spectra of the MFC in the $13^\circ$-twisted MC, where the driving field amplitude $h_1$ varies continuously from $5$ to $100 \, \text{mT}$ with all other parameters held constant. For $h_1 < 85 \, \text{mT}$, as $h_1$ increases, the amplitude of the incident mode at $\omega_1$ rises, and higher-order modes emerge progressively. In contrast, for $h_1 > 85 \, \text{mT}$, the amplitudes of the sum-frequency mode ($\omega_p / 2\pi = 10.5 \, \text{GHz}$) and the difference-frequency mode ($\omega_q / 2\pi = 9.5 \, \text{GHz}$) stop increasing and begin to decrease. Additionally, the higher-order modes become disordered and are overwhelmed by noise, as depicted in the bottom panel of Fig. \ref{Fig_h1}(b). This spectral disorder suggests that the system's ground state becomes unstable under such intense microwave excitation.
 
\begin{figure}[b]
    \centering
    \includegraphics[width=0.48\textwidth]{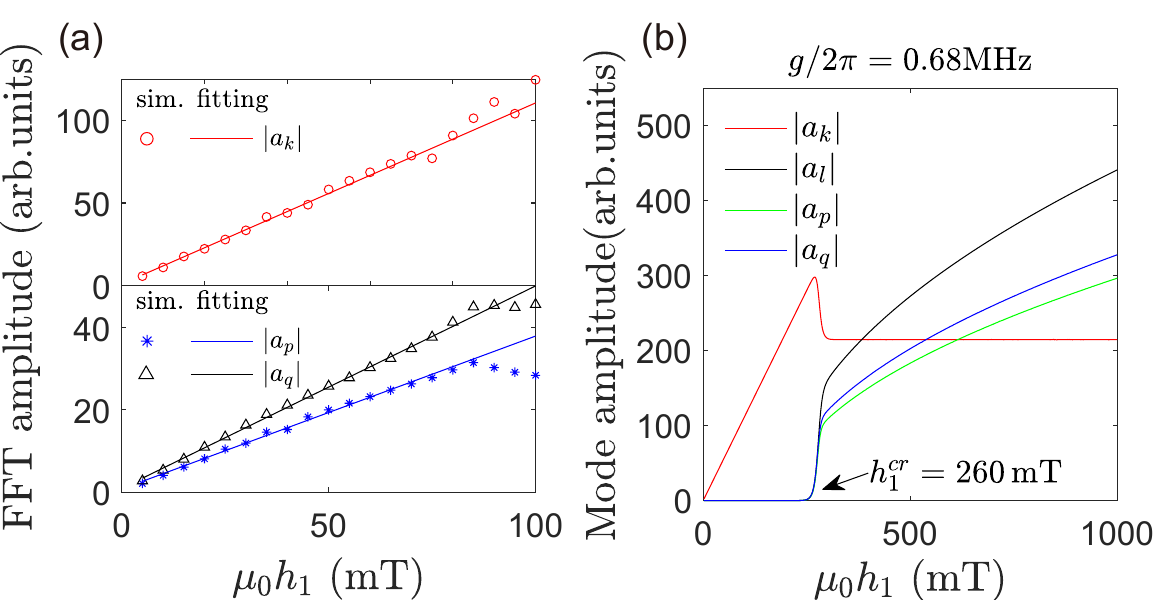}
    \caption{(a) Amplitudes of the three primary spectral peaks at $\omega_k$, $\omega_p=\omega_k + \omega_l$, and $\omega_q=\omega_k - \omega_l$ as a function of the driving field amplitude $h_1$ in the $13^\circ$-twisted MC, with $\omega_k / 2\pi=\omega_1 / 2\pi = 10 \, \text{GHz}$ and $\omega_l / 2\pi = 0.5 \, \text{GHz}$. (b) Theoretical curves of the mode amplitudes for the case where the second driving field is absent ($\mu_0 h_2 = 0$) and the nonlinear coupling strength is $g = 0.68 \, \text{MHz}$, illustrating the system's response under these conditions.}
    \label{Fig_threshold}
\end{figure}

The coupling strength $g_{p,q}$ between the incident magnon mode and the Kittel mode is a critical factor in the formation of the MFC. Analytical calculation of $g_{p,q}$ is challenging due to its dependence on the spatial overlap of the magnon mode profiles, which are complex in the system under study. To overcome this, we determine $g_{p,q}$ numerically by solving the Heisenberg equations of motion [Eqs. (\ref{Eq_Heisenberg})], and fitting the results to simulated data for driving field amplitudes below $\mu_0 h_1 = 85 \, \text{mT}$, as illustrated in Fig. \ref{Fig_threshold}(a). For simplicity, we assume $g_{p,q} = g$, a reasonable approximation given the comparable dynamics of the interacting modes in this context. The damping rate of the magnon modes is modeled as $\alpha_{k,l,p,q} = \alpha_0 + \delta \alpha$, where $\alpha_0 = 10^{-4}$ represents the intrinsic Gilbert damping constant used in the simulations, and $\delta \alpha = 0.02$ is an additional term introduced to account for losses due to absorbing boundary conditions and higher-order interactions not included in the simplified form of Eqs. (\ref{Eq_Heisenberg}). Additionally, the excitation efficiencies $\lambda_1$ and $\lambda_2$, which describe the coupling of the microwave fields to the magnon modes, are assumed to be equal ($\lambda_1 = \lambda_2 = \lambda$) and are determined through the fitting process to be $\lambda = 8.3 \gamma \mu_0$. The justification for the additional damping term $\delta \alpha$ and the detailed fitting procedures for $\delta \alpha$ and $\lambda$ are provided in Appendix \ref{section:fit_para}.

By numerically solving for the steady-state amplitudes of the magnon modes $|a_{k,l,p,q}|$ as functions of the driving field amplitude $h_1$ and the coupling strength $g$, we find that the ratio $|a_{p(q)}| / |a_k|$ is approximately proportional to $g$, provided that $g$ and $h_1$ remain within moderate limits (specifically, $\mu_0 h_1 < 85 \, \text{mT}$ and $g / 2\pi < 1 \, \text{MHz}$). This approximation enables us to determine the coupling strength $g$ through linear fitting of the simulated amplitude data, as presented in Fig. \ref{Fig_threshold}(a), which plots the amplitudes of the primary and secondary magnon modes against $h_1$. From this fitting, we obtain a coupling strength of $g / 2\pi = 0.68 \, \text{MHz}$.
To evaluate the feasibility of single-frequency driving, we substitute this value of $g$ into Eqs. (\ref{Eq_Heisenberg}) and set $h_2 = 0$ to calculate the threshold field $h_1^{\text{cr}}$ required to trigger nonlinear three-magnon processes when only $h_1$ is applied. This threshold, defined as the field strength at which the amplitudes of the secondary modes $a_p$ and $a_q$ become significant, is found to be $\mu_0 h_1^{\text{cr}} = 260 \, \text{mT}$, as depicted in Fig. \ref{Fig_threshold}(b). However, such a high driving field amplitude compromises the stability of the system's magnetic ground state, leading to chaotic spectral behavior and noise, as mentioned above. This explains why we adopt a two-tone driving approach, which leverages lower individual field amplitudes while still facilitating the necessary nonlinear interactions.

\begin{figure}[!t]
    \centering
    \includegraphics[width=0.48\textwidth]{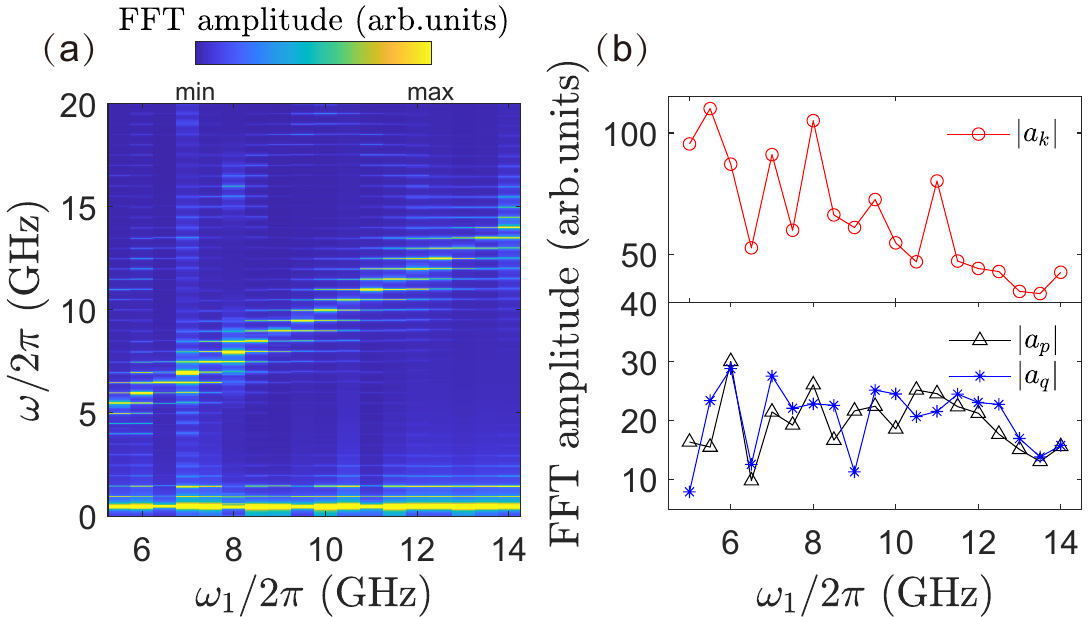}
    \caption{(a) Magnon spectra as a function of the driving frequency \(\omega_1\) for a twisted magnonic crystal with \(\theta = 13^\circ\), showing the evolution of the MFC across a broad frequency range. (b) Amplitudes of the primary spectral peaks at \(\omega_1\), \(\omega_1 + \omega_l\), and \(\omega_1 - \omega_l\) as functions of the driving frequency \(\omega_1\). }
    \label{Fig_omega1}
\end{figure}

We also investigate the influence of the driving frequency $\omega_1$ on the MFC. As illustrated in Fig. \ref{Fig_omega1}(a), the MFC is generated over a broad range of driving frequencies, demonstrating the system's versatility. As $\omega_1$ increases, the amplitude of the incident mode at $\omega_1$ generally decreases, consistent with the damping term $\alpha \omega_1$ in the Heisenberg equations of motion, Eqs. (\ref{Eq_Heisenberg}), though minor fluctuations are observed [see the top panel of Fig. \ref{Fig_omega1}(b)]. In contrast, the amplitudes of the sideband peaks at $\omega_1 \pm \omega_l$ oscillate around a relatively constant average value [see the bottom panel of Fig. \ref{Fig_omega1}(b)]. This behavior highlights the robustness of the MFC against variations in the driving frequency, unlike MFCs generated through nonlinear magnon-skyrmion scattering, which are typically confined to specific frequency windows due to the magnonic dispersion and spatial characteristics of the system's intrinsic modes \cite{Wang2021,Yao2023,Zhang2024}. 

\section{Conclusion}\label{section:D}
To conclude, we proposed a novel approach to generating MFCs in twisted MCs by harnessing three-wave mixing between the Kittel mode and a propagating magnon mode. This represents a notable advancement in magnonics, marking the first demonstration of MFC generation within a twisted magnetic system. To achieve this, we employed a two-tone microwave drive, which effectively overcomes the prohibitively high threshold associated with single-frequency excitation. This method not only reduces the required driving field strength but also enhances the stability of MFC generation. A central finding of our study is the significant enhancement of three-magnon processes at finite twist angles. This enhancement arises from the amplified non-collinearity of the static magnetization, driven by interlayer DDIs---a unique feature induced by the twisted structure. Under optimal conditions of twist angle and excitation frequency, we achieve high-quality MFCs with over 20 comb teeth, a performance level that surpasses many conventional magnonic platforms. Moreover, these MFCs exhibit remarkable robustness against variations in excitation frequency, distinguishing our approach from other methods, such as nonlinear magnon-skyrmion scattering, which are often constrained to narrow frequency ranges. This combination of tunability and resilience underscores the versatility of our system. Our study not only deepens the understanding of magnonic frequency combs but also introduces a versatile platform for future research and innovation. The synergy of twisted magnonic crystals, two-tone driving, and enhanced nonlinear interactions provides a powerful new tool for manipulating magnons, with far-reaching implications for both fundamental science and applied technologies in magnonics and beyond.
\begin{acknowledgments}
This work was funded by the National Key R\&D Program under Contract No. 2022YFA1402802, the National Natural Science Foundation of China (NSFC) (Grants No. 12374103 and No. 12434003) and Sichuan Science and Technology program (No. 2025NSFJQ0045).
\end{acknowledgments}

\begin{appendix}
\section{Two-Spin System}\label{section:two-spin_sys}
To illustrate how a non-collinear ground state enhances three-magnon processes, we analyze a system of two spins, $\mathbf{m}_1$ and $\mathbf{m}_2$, coupled by a ferromagnetic exchange interaction $J_{12}$. Anisotropy energy is included to manipulate the ground state (i.e., the static magnetization vectors). The Hamiltonian for this system is 
\begin{equation}\label{Eq_Hamiltonian_of_two_spins}
    {\mathcal H} = -J_{12}{\bf m}_1\cdot {\bf m}_2-K({\bf m}_1\cdot {\hat z}_1)^2-K({\bf m}_2\cdot {\hat z}_2)^2,
\end{equation}
where $J_{12}$ and $K$ are the exchange constant and anisotropy coefficient, respectively, and ${\hat z}_{1(2)}$ represent the direction of the easy axis. 

\begin{figure}[t]
    \centering
    \includegraphics[width=0.4\textwidth]{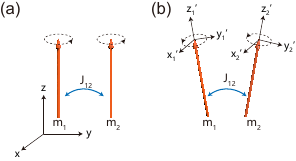}
    \caption{Schematic diagram of a system of two spins with (a) collinear and (b) non-collinear magnetizations \(\mathbf{m}_1\) and \(\mathbf{m}_2\), coupled by ferromagnetic exchange interaction \(J_{12}\).}
    \label{Fig_two_spin}
\end{figure}

For the collinear ground state as depicted in Fig. \ref{Fig_two_spin}(a), we take $\hat{z}_1=\hat{z}_2=(0,0,1)$, aligning both static magnetizations along the $z$-axis. To study magnon excitations, we apply the HP transformation to the normalized magnetization components, expressing them in terms of magnon creation ($a^\dagger$) and annihilation ($a$) operators
\begin{equation}
    \begin{aligned}
        m_x&=\frac{S^++S^-}{2S}\approx\frac{1}{\sqrt{2S}}(a+a^\dagger-\frac{a^\dagger aa+a^\dagger a^\dagger a}{4S}), \\
        m_y&=\frac{S^+-S^-}{2iS}\approx\frac{-i}{\sqrt{2S}}(a-a^\dagger-\frac{a^\dagger aa-a^\dagger a^\dagger a}{4S}),\\
        m_z&=1-\frac{a^\dagger a}{S},
    \end{aligned}
\end{equation}
where $S$ is the spin magnitude, $a$ and $a^\dagger$ obey the Bose commutation relation $[a_i, a_j^\dagger] = \delta_{i,j}$. Expanding the Hamiltonian in terms of $a$ and $a^\dagger$, we find only even-order terms (e.g., $a^\dagger a$), which conserve the magnon number. This indicates that three-magnon interactions, requiring odd-order terms (e.g., $a^\dagger a a$), are absent in the collinear case.
For the non-collinear ground state as depicted in Fig. \ref{Fig_two_spin}(b), we choose $\hat{z}_{1}=(0,-\sin\varphi,\cos\varphi)$ and $\hat{z}_{2}=(0,\sin\varphi,\cos\varphi)$ for simplicity. 
Minimizing the total energy yields the static magnetizations ${\bf m}_1=(0,-\sin\theta,\cos\theta)$ and ${\bf m}_2=(0,\sin\theta,\cos\theta)$ with $\theta=\frac12\arctan{(\frac{K\sin{2\varphi}}{J+K\cos{2\varphi}})}$. The global coordinate $(x,y,z)$ needs be transformed to local coordinates $(x_{1(2)}',y_{1(2)}',z_{1(2)}')$ by rotation matrices, with $\hat {z}_{1(2)}'={\bf m}_{1(2)}$,
\begin{equation}
    \begin{pmatrix}
        m_{1(2)x} \\m_{1(2)y} \\m_{1(2)z} \\
    \end{pmatrix}
    =
    \begin{pmatrix}
        1 & 0 & 0 \\
        0 & \cos\theta & \mp\sin\theta \\
        0 & \pm\sin\theta & \cos\theta \\
    \end{pmatrix}
    \begin{pmatrix}
        m_{1(2)x'} \\m_{1(2)y'} \\m_{1(2)z'} \\
    \end{pmatrix}.
\end{equation}
Consequently the third-order nonlinear interactions induced by exchange energy emerge from the product of $m_{1(2)y'}$ and $m_{1(2)z'}$ and is proportional to $\sin2\theta$, 
\begin{equation}
    \begin{aligned}
        \mathcal{H}_{ex}^{(3)}/J_{12} &= \frac{i}{\sqrt{2}S^{3/2}}\sin2\theta [-a_1a_2^\dagger a_2+a_1^\dagger a_2^\dagger a_2+a_1^\dagger a_1a_2-a_1^\dagger a_1a_2^\dagger \\
        &-\frac{1}{4}(a_1^\dagger a_1a_1-a_1^\dagger a_1^\dagger a_1-a_2^\dagger a_2a_2+a_2^\dagger a_2^\dagger a_2)].
    \end{aligned}
\end{equation}
This third-order Hamiltonian is proportional to $\sin 2\theta$, which is zero in the collinear case ($\theta = 0$) but nonzero when the magnetizations diverge ($\theta \neq 0$). Thus, non-collinearity enables three-magnon interactions, such as magnon splitting or confluence, which are absent in the collinear ground state.

This two-spin model demonstrates that a non-collinear ground state, induced by misaligned anisotropy axes, activates three-magnon processes through third-order terms in the Hamiltonian. This mechanism is a simplified analog of effects observed in twisted magnonic crystals, where non-collinearity enhances nonlinear magnon dynamics.

\section{Fitting Procedure of $\delta\alpha$ and $\lambda$}\label{section:fit_para}

\begin{figure}[!b]
    \centering
    \includegraphics[width=0.48\textwidth]{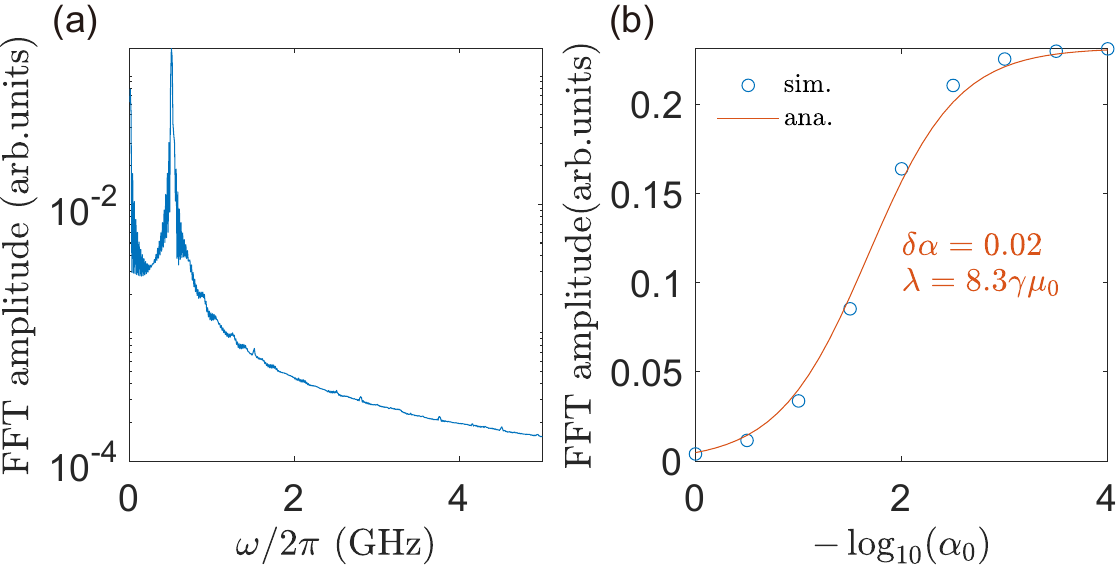}
    \caption{(a) The magnon spectrum when $\mu_0h_1=0$, $\mu_0h_2=0.01\,\rm mT$ and $\alpha_0=10^{-4}$. (b) The simulated and analytical mode amplitude as a function of $\alpha_0$.}
    \label{Fig_fitting_paras}
\end{figure}

To account for the effects of absorbing boundary conditions and higher-order interactions not included in the simplified model of Eq. (\ref{Eq_Heisenberg}), we introduce an additional damping term $\delta \alpha$. This term modifies the effective damping constant to $\alpha = \alpha_0 + \delta \alpha$, where $\alpha_0$ represents the intrinsic Gilbert damping constant.
We investigate this by applying a single-frequency microwave field to drive the system, with a frequency of $\omega / 2\pi = 0.5 \, \rm GHz$ and a small amplitude of $\mu_0 h = 0.01 \, \rm mT$. Under these conditions, nonlinear interactions are negligible, and only magnons at the driving frequency $\omega$ are excited. This is illustrated in Fig. \ref{Fig_fitting_paras}(a), which displays the spectrum of excited magnons. Starting from the Heisenberg equations of motion, Eq. (\ref{Eq_Heisenberg}), we derive the steady-state amplitude of the magnon mode at frequency $\omega$. For a single-frequency drive, we set the second driving field to zero ($\mu_0 h_1 = 0$), neglect nonlinear coupling ($g = 0$), and assign $h_2 = h$ and $\omega_2 = \omega$. This yields
\begin{equation}\label{Eq_fitting_alpha}
    |a|=\frac{\lambda h}{\alpha\omega},\alpha=\alpha_0+\delta \alpha. 
\end{equation} 

To determine the additional damping $\delta \alpha$
 and the excitation efficiency $\lambda$, we perform micromagnetic simulations. In these simulations, we vary the intrinsic damping constant $\alpha_0$ from $10^{-4}$ to $1$ and compute the corresponding magnon amplitudes $|a|$. By fitting the simulated amplitudes to the expression above, we obtain $\delta \alpha = 0.02$ and $\lambda = 8.3 \gamma \mu_0$.
The simulated data points and the resulting fitting curve are presented in Fig. \ref{Fig_fitting_paras}(b). The excellent agreement between them validates our assumption that the additional damping $\delta \alpha$ effectively accounts for the effects of absorbing boundary conditions and higher-order interactions.
\end{appendix}

\end{document}